\documentclass[aps,pra,superscriptaddress,notitlepage,twocolumn]{revtex4-1}
\usepackage[utf8]{inputenc}
\usepackage{braket}
\usepackage{amsmath,amsfonts}
\usepackage{graphicx}
\usepackage{bm}
\usepackage[colorlinks]{hyperref}

\newcommand{\ketbra}[2]{\ket{#1} \hspace{-0.8ex}\bra{#2}}
\newcommand{\nty}[1]{\mathcal{N}\left[ #1 \right]}
\newcommand{\norm}[1]{\| #1 \|}
\newcommand{\Tr}{\mathrm{Tr}}

\newcommand{\bbC}{\mathbb{C}}

\newcommand{\ba}{\begin{eqnarray}}
\newcommand{\ea}{\end{eqnarray}}
\newcommand{\ban}{\begin{eqnarray*}}
\newcommand{\ean}{\end{eqnarray*}}
\newcommand{\Id}{\openone}
\newcommand{\balpha}{{\bm{\alpha}}}
\newcommand{\bx}{{\bm{x}}}
\newcommand{\by}{{\bm{y}}}

\newcommand{\change}[1]{{#1}}
\newcommand{\pending}[1]{{#1}}

\begin{document}

\author{Yu Cai}
\affiliation{Department of Applied Physics, University of Geneva, Geneva, Switzerland}

\author{Baichu Yu}
\affiliation{Centre for Quantum Technologies, National University of Singapore, 3 Science Drive 2, Singapore 117543, Singapore}

\author{Pooja Jayachandran}
\affiliation{Centre for Quantum Technologies, National University of Singapore, 3 Science Drive 2, Singapore 117543, Singapore}

\author{Nicolas Brunner}
\affiliation{Department of Applied Physics, University of Geneva, Geneva, Switzerland}

\author{Valerio Scarani}
\affiliation{Centre for Quantum Technologies, National University of Singapore, 3 Science Drive 2, Singapore 117543, Singapore}
\affiliation{Department of Physics, National University of Singapore, 2 Science Drive 3, Singapore 117542, Singapore}

\author{Jean-Daniel Bancal}
\affiliation{Department of Applied Physics, University of Geneva, Geneva, Switzerland}

\begin{abstract}
    The notion of entanglement of quantum states is usually defined with respect to a fixed bipartition. Indeed, a global basis change can always map an entangled state to a separable one. The situation is however different when considering a set of states. In this work we define the notion of an ``absolutely entangled set'' of quantum states: for any possible choice of global basis, at least one of the states in the set is entangled. Hence, for all bipartitions, i.e. any possible definition of the subsystems, the set features entanglement. We present examples of such sets, including sets with minimal size. Moreover, we propose a quantitative measure for absolute set entanglement. To lower-bound this quantity, we develop a method based on polynomial optimization to perform convex optimization over unitaries, which is of independent interest.
\end{abstract}

\title{Entanglement for any definition of two subsystems}
\maketitle

\textit{Introduction.---} Composite quantum systems can be found to be in \textit{entangled states}, a direct consequence of the linearity of quantum mechanics. This concept has far-reaching implications, and is by now considered as one of the defining features of quantum theory \cite{Horodecki_RMP,Guhne_Toth}. 

The notion of entanglement relies on \textit{partitioning the system into subsystems}. Some choice may be very natural to us, notably the one based on localization: what is accessible in Alice's location, versus what is accessible in Bob's location. In other cases, the arbitrariness of the partition is more patent. Consider a two-path interference for a molecule. If one chooses the $N$ atoms that form the molecule as subsystems, the state of the molecule in the interferometer will be a highly entangled state of the Greenberger-Horne-Zeilinger type: $\bigotimes_{j=1}^N\ket{\vec{x}_j+\vec{d}_{I}}+\bigotimes_{j=1}^N\ket{\vec{x}_j+\vec{d}_{II}}$, where $\vec{x}_j$ is the position of the $j$-th atom. But if one chooses the centre-of-mass (CM) and the relative coordinates as subsystems, the state of the molecule will be product: $(\ket{\vec{d}_I}+\ket{\vec{d}_{II}})_{CM}\otimes\ket{\vec{r}_1}\otimes...\otimes\ket{\vec{r}_{N-1}}$, where $\vec{r}_k$ are the relative coordinates. Philosophers may discuss whether some choices of subsystems represent ``reality'' better than others \cite{KJJ,earman}. In the practice of the physicist, the definition of subsystems relies on operational convenience \cite{zanardi2001,zanardi2004,harshman2011}. To stay with the example (see also \cite{JAD}): we may well believe that molecules are ``really'' made of atoms; this division may also prove suited for calculations in quantum chemistry; but the description of CM and relative coordinates is more convenient to describe path interference.

For another related example, consider the second quantisation of bosons. The Fock space is constructed as a tensor product of spaces, each representing a bosonic mode. A change of mode decomposition reads $A_j=\mathcal{U}a_j=Ua_jU^\dagger$, where $\mathcal{U}$ is a unitary operator acting on the modes and $U$ is the corresponding representation in the Hilbert space. Any $\mathcal{U}$ is allowed, but the corresponding $U$ must be a linear-optics transformation. Thus, if one assumes that the only meaningful tensor decompositions of the field are those in modes, not all quantum states can be connected. For instance, the entangled state $\frac{1}{\sqrt{2}}(\ket{2,0}+\ket{0,2})=\frac{1}{2}({a_1^\dagger}^2+{a_2^\dagger}^2)\ket{\textrm{vacuum}}$ can be written as product by changing modes \cite{vanenk2003}, since it is equal to $\ket{1,1}=A_+^\dagger A_{-}^{\dagger}\ket{\textrm{vacuum}}$ for $A_{\pm}=\frac{1}{\sqrt{2}}(a_1\pm ia_2)$. But there is no mode transformation in which the same state can be written $\frac{1}{\sqrt{2}}{A^{\dagger}}^2\ket{\textrm{vacuum}}=\ket{2,0}$; not to mention the impossibility of connecting states with different number of photons. In this context, some states were recently found that remain entangled under any mode transformation \cite{sperling2019}.

Such results are clearly impossible if one considers all the unitaries $U$ in the Hilbert space, as we plan to do here (in the context of distinguishable systems). In this case, given $\ket{\psi}$, there exists $U$ such that $U\ket{\psi}=\ket{\phi}$ for any target state $\ket{\phi}$. However, the situation becomes completely different when one considers \textit{sets of quantum states}, as we discuss in this work. Indeed, there  exist sets of quantum states, from which the entanglement cannot be removed even by global unitaries. That is, entanglement will remain no matter what definition of the subsystems is adopted. We term such sets ``absolutely entangled''. \change{We formally define this notion, followed by a warm-up discussion. We prove a general lower bound on the size of an absolutely entangled set: for $\bbC^{d}=\bbC^{d_1}\otimes\bbC^{d_2}$, one needs a set of at least $\max{(d_1,d_2)}+2$ pure states, while for any smaller set there is a choice of global basis such that all states become product. Then we present an explicit construction of an absolutely entangled set consisting of $d_1+d_2$; this set is therefore minimal if $\min{(d_1,d_2)}=2$.} Further, we quantify the amount of entanglement present in an absolutely entangled set. For this, we develop a method for performing convex optimization over unitaries, which can be cast as a semidefinite programming (SDP). We present several illustrative case studies. Finally, we discuss the potential applications of these ideas, as well as some open questions.

\textit{Context and definition.---} As mentioned above, a single quantum pure state is always unitarily equivalent to a product state. Similarly, for any mixed quantum state $\rho$, which can always be expressed as a probabilistic mixture of pure states forming an orthonormal basis $\{\ket{\psi_j}\}$, there exists a unitary $U$ that maps $\{\ket{\psi_j}\}$ to the computational basis (containing only product states), and thus maps $\rho$ to a separable state. Hence a single quantum state, whether pure or mixed, can only be entangled with respect to some bipartitions \cite{Thirring_2011}. Note that this is in fact not even always the case: there exist mixed states (in particular in the vicinity of the maximally mixed state) that remain separable for any global basis choice \cite{Johnston_2013}.

\change{The situation becomes completely different if one consider sets of states. Indeed, there exist sets of states, for which any global basis choice will leave at least one state in the set entangled. A trivial example is the set of \textit{all pure states} in a given Hilbert space $\bbC^{d_1} \otimes \bbC^{d_2}$: clearly no unitary can map all those states into product ones \footnote{Given the vast body of work on quantum entanglement, it is not surprising that absolutely entangled sets of states have been chanced upon while investigating other properties. Concretely, in the study of mutually unbiased bases (MUBs), it was noticed early on that the five MUBs in $\bbC^4$ constitute a set of 20 pure states, only 12 of which (3 bases) can be made product when seen in $\bbC^2\otimes\bbC^2$ \cite{RBKS05}. The maximal number of MUBs that can be product was later given for any number of parties of arbitrary dimensions \cite{WPZ11}.}.} This motivates the following:

\change{{\bf Definition: absolutely entangled set (AES).} \textit{Consider a set of quantum states $\{\rho_1,...,\rho_K\}$ in a fixed Hilbert space $\bbC^d$ of non-prime dimension. The set is said to be absolutely entangled with respect to all bipartitions into subsystems of dimension $(d_1,d_2)$, if for every unitary $U \in SU(d)$, at least one state $U\rho_kU^\dagger$ is entangled with respect to $\bbC^{d_1} \otimes \bbC^{d_2}$.}

When $d$ is not a product of primes, one could consider a stronger definition, by requesting that at least one state remains entangled for all $d_1,d_2\geq 2$ such that $  \bbC^{d}= \bbC^{d_1} \otimes \bbC^{d_2}$, rather than for a fixed pair of dimensions; but we shall leave this stronger definition for further work.}


Let us discuss a warm-up example in $\bbC^4$. As noted above, any orthonormal basis (such as the Bell basis) is unitarily equivalent to the computational basis. Then, a natural AES candidate that comes to mind (at least for the authors) is the set in $\bbC^4$ consisting of the computational basis and the four Bell states: $\{ \ket{00}, \ket{01}, \ket{10}, \ket{11}, \ket{\Phi^+}, \ket{\Phi^-}, \ket{\Psi^+}, \ket{\Psi^-}\}$. But this set is easily dismissed: a standard CNOT unitary transforms all those states into products, since $U_{\textrm{CNOT}}\ket{\Phi^+}= \ket{+}\ket{0}$ and so on. By the same token, we see immediately that any set of states \textit{can be made product} by a suitable global unitary (a CNOT), whenever these states are \textit{all Schmidt-diagonal in the same computational basis}, i.e.~if some can be written as $\cos\theta_j\ket{00}+\sin\theta_j\ket{11}$ and the others as $\cos\theta'_j\ket{01}+\sin\theta'_j\ket{10}$.

\textit{Lower bound on the number of states.---} \change{For any bipartition $\bbC^d=\bbC^{d_1} \otimes \bbC^{d_2}$, we are going to show that no set of $\max{(d_1,d_2)}+1\equiv d'+1$, or fewer, pure states is an AES. Indeed,} without loss of generality, one can always write the first $d'+1$ states as:
\begin{align}
\begin{split}
    \ket{\psi_1} & = \ket{0}\ket{0}, \\
    \ket{\psi_2} & = (c_{2,0} \ket{0} + c_{2,1} \ket{1})\ket{0}, \\
    \ket{\psi_3} & = (c_{3,0} \ket{0} + c_{3,1} \ket{1} + c_{3,2} \ket{2})\ket{0},\\
    & \vdots \\
    \ket{\psi_{d'}} & = \left(\sum_{i=0}^{d'-1} c_{d',i} \ket{i}\right) \ket{0}.
\end{split}
\end{align}
The coefficient are determined by the Gram matrix $\braket{\psi_i|\psi_j}$, for example $c_{2,0} = \braket{\psi_1|\psi_2}$ and $c^*_{2,0}c_{3,0}+c_{2,1}^*c_{3,1} = \braket{\psi_2|\psi_3}$. Now, the overlap of the $(d'+1)$-th state with the previous ones will be fully encoded in the component with the second system in state $\ket{0}$: we are therefore completely free to choose how to write the component with the second system in an orthogonal state $\ket{1}$. In particular, we can write 
\begin{align}
\begin{split}
    \ket{\psi_{d+1}} & = \left(\sum_{i=0}^{d-1} c_{d+1,i} \ket{i} \right) \ket{0} + c_{d+1,d+1}\left(\sum_{i=0}^{d-1} c_{d+1,i} \ket{i} \right) \ket{1} \\
    & = \left(\sum_{i=0}^{d-1} c_{d+1,i} \ket{i} \right) (\ket{0} + c_{d+1,d+1}\ket{1}).
\end{split}
\end{align}
Thus, there exists a basis in which all the $d'+1$ states are product. \change{It is natural to ask whether this construction is tight, i.e.~if one can find absolutely entangled sets of $d'+2$ states. The next explicit construction proves that this is the case for $\min{(d_1,d_2)}=2$ (i.e. for $d=2d')$. Tighteness for $\min{(d_1,d_2)}>2$ remains an open question, as well as proving that the lower bound on the size of a AES remains valid for mixed states.}

\textit{Minimal absolutely entangled sets.---} \change{Our first explicit construction of an AES is as follows. Consider $d$ non-prime and the bipartition $\bbC^d=\bbC^{d_1} \otimes \bbC^{d_2}$, and let $\{\ket{\xi_i}\}_{i=1,...,d}$ be an orthonormal basis of $\bbC^d$. The $K\equiv d_1+d_2$ states
\ba\label{set4}
\begin{array}{ccl}
    \ket{\phi_1} &=& \ket{\xi_1}, \\
    \ket{\phi_k} &=& c\ket{\xi_{1}}+\sqrt{1-c^2}\ket{\xi_{k}},\;k=2,...,K
    \end{array}
\ea
form an AES when $c\in(\sqrt{\frac{(d_1-1)(d_2-1)}{d_1d_2}},1)$. In particular, for $d=2d'$ and the partition $d_1=d'$ and $d_2=2$, this set consists of $K=d'+2$ states. In view of the previous result, this is the \textit{minimal} size of a AES of pure states in that partition. Notice also that in general $K\leq d$, with equality if and only if $d=4$. 

The proof for arbitrary $(d_1,d_2)$ is given in Appendix~\ref{app:proof}. Here we give a proof for $d=4$ (i.e.~$d_1=d_2=2$) that actually applies to a larger family: for $k=2,3,4$, we allow $\ket{\phi_k}=c_{k1}\ket{\xi_{1}}+c_{kk}\ket{\xi_{k}}$ with possibly different coefficients $c_{k1}$.} Without loss of generality, let's consider a global $U$ such that $U\ket{\phi_1}$ is product and denoted $U\ket{\phi_1}=U\ket{\xi_1}=\ket{00}$. Thus, for $k=2,3,4$, we have $U\ket{\xi_k}=b_{k2}\ket{01} +b_{k3}\ket{10}+b_{k4}\ket{11}$. So
\ba
U\ket{\phi_k}&=&c_{k1}\ket{00}+c_{kk}(b_{k2}\ket{01} +b_{k3}\ket{10}+b_{k4}\ket{11})
\ea
and we want to see if these three states can all be made product too, which is the case if and only if \ba
c_{k1}b_{k4}=c_{kk}b_{k2}b_{k3}\label{eqproof}\ea for $k=2,3,4$ (notice that $c_{kk}\neq 0$, otherwise $\ket{\phi_k}=\ket{\phi_1}$ and the problem becomes trivial). Let us now impose $|c_{k1}|\in (\frac{1}{2},1)$ for all $k=2,3,4$: it follows from \eqref{set4} that $0<|c_{kk}/c_{k1}|< \sqrt{3}$, which inserted into \eqref{eqproof} gives $|b_{k4}|<\sqrt{3}|b_{k2}b_{k3}|$. But by normalization, $|b_{k2}b_{k3}|\leq\frac{1}{2}(1-|b_{k4}|^2)$. So we have found that a necessary condition for $U\ket{\phi_k}$ to be product is $|b_{k4}|<\frac{1}{\sqrt{3}}$. We have not yet used the fact that the three $U\ket{\xi_k}$ must be orthogonal. That condition implies that it is impossible for all three $|b_{k4}|$ to be strictly smaller than $\frac{1}{\sqrt{3}}$ \footnote{The proof is straightforward: if the vectors $U\ket{\xi_k}$ are orthogonal, the $3\times 3$ matrix $W$ whose coefficients are $w_{ij}=b_{i-1,j-1}$ is unitary. Thus, the square of the elements of each row and column must sum up to 1. In particular, the column $j=3$ must be such that $|b_{24}|^2+|b_{34}|^2+|b_{44}|^2=1$}. Thus, it is impossible to make all three $U\ket{\phi_k}$ product. As soon as one of the $|c_{k1}|\leq\frac{1}{2}$, there are instances where one can make all four states in \eqref{set4} separable (see Appendix~\ref{sec:sep}).


\textit{Quantitative approach: absolute set negativity---} More features of AESs can be uncovered by a quantitative approach to the \textit{minimal amount of entanglement} present in a set of states. Let again $\{ \rho_k\}_{k=1,..., K}$ be a set of $K$ states in dimension $d=d_1d_2$. The \textit{absolute set entanglement} with respect to all bipartitions $\bbC^d=\bbC^{d_1}\otimes\bbC^{d_2}$ can be then quantified as
\begin{align}\label{eq:Etot}
    E_{(d_1,d_2)}[{\{ \rho_k \}_k}] = \min_{U \in SU(d)} \sum_k E_{(d_1,d_2)}(U\rho_k U^\dagger)
\end{align}
where $E_{(d_1,d_2)}$ is an entanglement measure \cite{Horodecki_RMP,Guhne_Toth}. This figure of merit can be understood as follows: if a user receives states from a source that samples uniformly from the set $\{ \rho_k\}_{k=1,..., K}$, the average amount of entanglement he receives per round is at least $E_{(d_1,d_2)}[\{\rho_k\}_k]/K$. In what follows, we drop the subscript $(d_1,d_2)$.

As an entanglement measure, we choose \textit{negativity}~\cite{negativity} (in Appendix \ref{sec:entropy} we consider also the entropy of entanglement for some examples of sets of pure states). The negativity of a quantum state $\rho$ is given by
$\nty{\rho} = \frac{\norm{\rho^{T_A}}_1-1}{2} = \sum_{\lambda_i<0} | \lambda_i |$ where $\lambda_i$ are the eigenvalues of $\rho^{T_A}$, and $T_A$ refers to partial transposition on subsystem $A$. Alternatively, the negativity can be expressed through its variational definition as $ \nty{\rho} = \{\min_{\sigma^{\pm}} \Tr[\sigma^-]\,
    \vert\, \sigma^{\pm} \text{ are PPT}, \;
    \rho = \sigma^+ -  \sigma^- \}$, where PPT stands for positive partial transpose.
This allows one to express the absolute set negativity of $\{ \rho_k\}$ as
\begin{equation}
\begin{split}\label{eq:unitaryOpt}
    \nty{ \{ \rho_k \} } = \min_{U,\sigma^{\pm}_k} & \quad \sum_k \Tr[\sigma^-_k] \\
    \text{s.t.} & \quad \sigma^{\pm}_k \text{ are PPT}, \\
    & \quad U \rho_k U^\dagger= \sigma^+_k -  \sigma^-_k, \\
    & \quad U U^\dagger = U^\dagger U = \Id.
\end{split}
\end{equation}
Note that this optimization is not an SDP, because the second and third constraints are quadratic in $U$. To solve this problem, we now introduce a method to relax unitary constraints based on polynomial optimization, which allows us to cast the above problem as a family of semi-definite programs. In turn, we generalize the notion of localizing matrices to semi-definite constraints~\cite{Pironio2010,lasserre2001global}.

\textit{Convex optimization over unitaries ---}
Since unitary operators are defined by a quadratic constraint $U U^\dag=\mathbb{I}$, unitary optimization is nonlinear. Moreover, the set of unitaries is not convex: typically, $(U_1+U_2)/2$ is not a unitary. This makes unitary optimization particularly non-trivial in general.

In some cases, unitary optimization can be greatly simplified though. For instance, when considering such optimization with a linear (or concave) objective function, the non-convexity can be avoided by considering a simple (tight) relaxation of the problem, namely the optimization of the same objective function over the set of convex mixtures of unitaries. Since this set is convex by assumption, the optimal value for both optimizations coincides. Moreover, optimizing over convex mixtures of unitaries can be achieved easily by writing the problem in the Choi formalism, in which case the nonlinearity amounts to a semi-definite constraint which can be described efficiently (see for instance Eq.(39) and Supplementary Information B.3.1 of~\cite{branciard2016witnesses}).

In the case of Eq.~\eqref{eq:unitaryOpt}, however, we cannot benefit from this simplification: a mixture of all two-qubit unitaries constitute a depolarizing channel which leave no state entangled, and therefore relaxing Eq.~\eqref{eq:unitaryOpt} to allow optimization over mixture of unitaries only provides the trivial lower bound 0. In order to obtain a strict and nontrivial lower bound, we reformulate our optimization as a particular case of polynomial optimization~\cite{lasserre2001global}. For this, we parametrize $U$ as a $d \times d$ complex matrix with components $u_{i,j}\in\bbC$. The unitarity conditions then correspond to quadratic constraints on the components $u_{i,j}$:
\begin{equation}
\begin{split}\label{eq:unitaryOptElement}
    \nty{ \{ \rho_k \} } = \min_{u_{i,j},\sigma^{\pm}_k} & \quad \sum_k \Tr[\sigma^-_k] \\
    \text{s.t.} & \quad \sigma^{\pm}_k \text{ are PPT}, \\
    & \quad \sum_{l,m} u_{i,l} (\rho_k)_{l,m} u^*_{j,m}= (\sigma^+_k -  \sigma^-_k)_{i,j}, \\
    & \quad \sum_m u^*_{m,i} u_{m,j} = \sum_m u_{i,m} u_{j,m}^* = \delta_{i,j}
\end{split}
\end{equation}

At this stage it is clear that applying the SDP relaxation method of~\cite{lasserre2001global} to the polynomial variables $u_{i,j}$ allows one to obtain a hierarchy of SDP that captures the unitarity part of Eq.~\eqref{eq:unitaryOpt}. Introducing additionally a notion of localizing matrices for semidefinite constraints involving the matrix variables $\sigma_k^\pm$ allows us to formulate our problem~\eqref{eq:unitaryOpt} fully in terms of semi-definite programming (see Appendix~\ref{sec:convex} for full details, \change{including the localizing matrix for each constraint used to obtain the results in the case studies}).

\textit{Case studies
---} The above SDP relaxation method allows one to obtain \change{a certifiable} lower-bound on the absolute set negativity. On the other hand, an upper bound can be computed by heuristic numerical minimization (here we use \texttt{fminunc} in Matlab).

\begin{figure}[tb]
    \centering
    \includegraphics[scale=0.8, trim=0cm 1cm 0cm 1cm, clip]{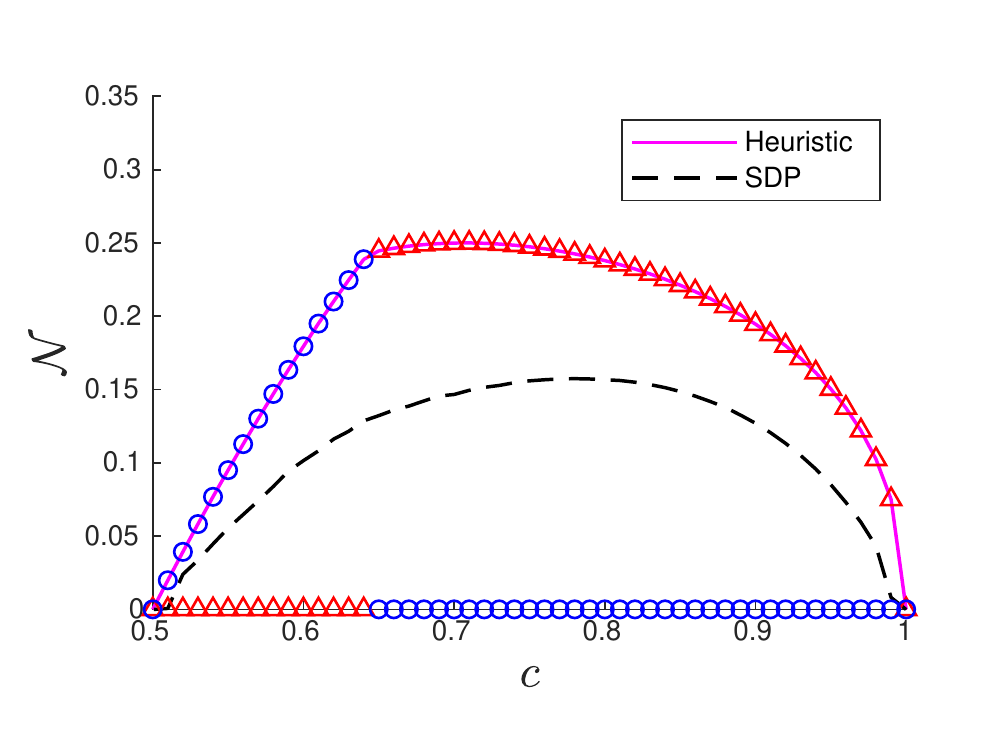}
    \caption{Absolute set negativity $\mathcal{N}$, as defined in Eq.\eqref{eq:unitaryOpt}, for the one parameter set of states Eqs.\eqref{set4} for $d=4$. The heuristic upper bound (magenta line) is computed by \texttt{fminunc}; the lower bound (black dashed) is computed by the SDP relaxation. For the heuristic minimization, we plot also the negativity in the first state (blue circles) and in one of the other three (red triangles).} 
    \label{fig:minNeg}
\end{figure}

We study first the one-parameter family of states \eqref{set4}. The results are plotted in Fig.~\ref{fig:minNeg}. The gap between the upper and lower bounds suggests that the SDP method does not give tight bounds \change{(we expect the heuristic method to given values which are close to optimal, as optimisation involves here 15 real parameters for parametrizing $U\in SU(4)$)}: nonetheless, \change{the SDP method} does provide a nontrivial \change{and certifiable} lower bound, i.e. $\mathcal{N}>0$, for the entire range $c \in (\frac{1}{2},1)$ covered by the analytical proof. Moreover, in the heuristic optimization one can check how negativity is distributed among the states. By inspection, it turns out that it is better to concentrate all the negativity in one state. For $c\lesssim 0.65$, this is the first state (which has scalar product equal to $c$ with the three other states); for $c\gtrsim 0.65$, it is one of the other three symmetric states.

Next, we look for the set of four state featuring the largest absolute set negativity. Via a heuristic see-saw algorithm we found the following candidate:
\begin{align}\label{set2}
    \begin{split}
        \ket{\varphi_1} &= \ket{\xi_1}, \\
        \ket{\varphi_2} &= a\ket{\xi_{1}}+b\ket{\xi_{2}}+b\ket{\xi_{3}}+c\ket{\xi_{4}}, \\
        \ket{\varphi_3} &= a\ket{\xi_{1}}+b\ket{\xi_{2}}+c\ket{\xi_{3}}+b\ket{\xi_{4}}, \\
        \ket{\varphi_4} &= a\ket{\xi_{1}}+c\ket{\xi_{2}}+b\ket{\xi_{3}}+b\ket{\xi_{4}},
    \end{split}
\end{align}
where $a=0.6245$,\\ $b=\frac{1}{3} \sqrt{-3a^2+a -2(1-a)\sqrt{3a+1}+2}$ and $c = \sqrt{1-a^2-2b^2}$. For this set, the SDP lower bound is $\mathcal{N} = 0.2213$ and the heuristic upper bound is $\mathcal{N} = 0.4609$, both clearly exceeding the maximal values for the previous case study shown in Fig.~\ref{fig:minNeg}.

\begin{figure}[tb]
    \centering
    \includegraphics[scale = 0.8, trim=0cm 1cm 0cm 1cm, clip]{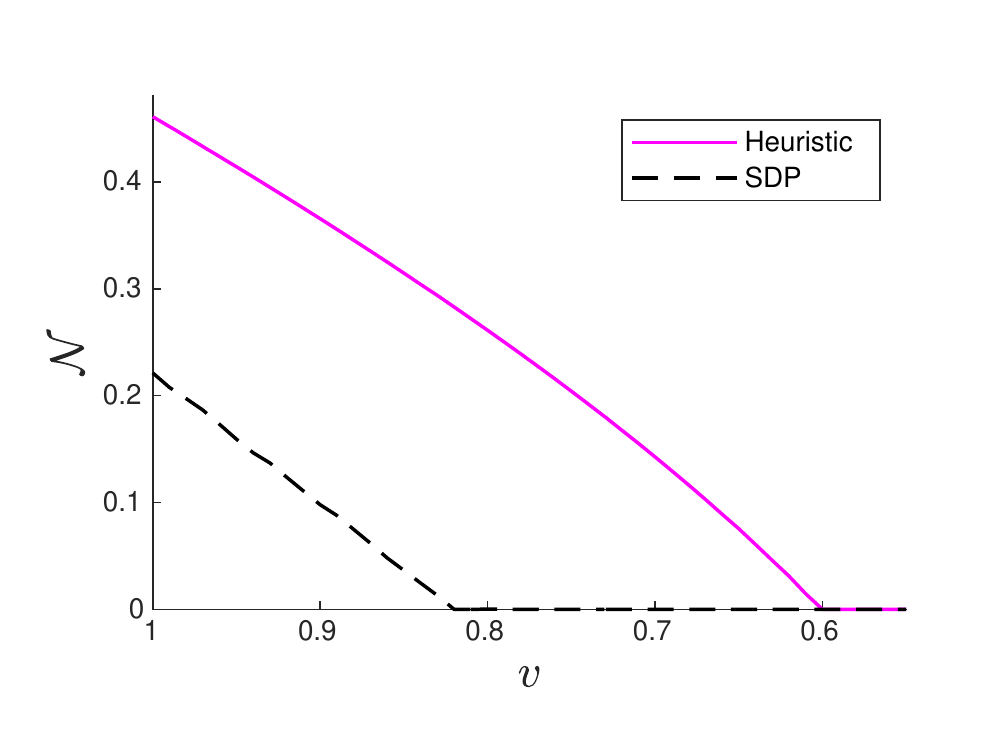}
    \caption{Absolute set negativity $\mathcal{N}$ for the set of four states in Eqs.\eqref{set2} mixed with an amount $1-v$ of white noise. From heuristic optimization, it appears that the set for $v=1$ features the largest absolute set negativity in $\bbC^4$. The curves are the upper and lower bounds as in Fig.~\ref{fig:minNeg}.}
    \label{fig:noisyMES}
\end{figure}

We start from the same family of states to extend the definition of AESs to mixed states and show that absolute entanglement is robust to noise. Specifically, we consider the set of mixed states $\rho_k = v\ketbra{\varphi_k}{\varphi_k}+(1-v)\frac{\Id}{4}$, where each pure state in \eqref{set2} is mixed with a given amount of white noise. The upper and lower bound on $\mathcal{N}$ are shown in Fig.~\ref{fig:noisyMES}, as a function of the visibility $v$. Entanglement vanishes below a critical visibility, whose value is between $0.82$ (SDP) and $0.6$ (heuristic).

\textit{Towards an operational application.---} Our first result, namely, that $d+1$ states in $\bbC^{2d}$ can always be made all product, can be rephrased as: any set of $n$ states can always be transformed into all product states, if they are seen as states in $\bbC^{2(n-1)}$. This observation may be given a polemic twist: if we are ready to redefine subsystems, shouldn't we also question the identification of the relevant degrees of freedom, i.e.~the identification of the total system, at least in principle? Leaving again metaphysics aside, one may answer that the reductionism involved in identifying a system is a necessary step of the scientific method. But when quantum entanglement is involved, \change{more is possible, specifically in the situations of self-testing (see \cite{SB19} for a recent review). Device-independent self-testing relies on no-signaling, that is, on having spatially separated subsystems; but one of the possible deductions is that inside \textit{one} of the black boxes there is a composite system \cite{rabelo2011,RUV}. The initial necessity of spatially separated subsystems can be replaced by computational assumptions on the verifier \cite{MV20}. Thus, with these tools, a ``total system under study'' can be identified operationally within one black-box, including the fact that it must be composite in character. Further certifying that there is an entangled state inside one black box was proved feasible in very special setups \cite{rabelo2011,bancal2015}. Our notion of AESs provides a much greater freedom in designing such certifications. It may also help in addressing some open problems, notably finding a robust bound for the device-independent certification of ``irreducible dimension''\cite{cong2017witnessing}.}

\textit{Conclusion.---} We have introduced a notion of absolutely entangled sets (AESs) of quantum states, i.e. featuring entanglement for any possible definition of two subsystems. \change{We have given several explicit examples of such sets of pure states, that are provably minimal for partitions $\bbC^d=\bbC^{d'}\otimes\bbC^2$. We also developed a quantitative approach to this phenomenon: the minimal entanglement present in a set of states was upper bounded by heuristic optimisation, and lower bounded by a new method of convex optimization over unitaries, also generalizing the concept of localizing matrices to semi-definite constraints (the latter method may be of independent interest and apply to a broader range of problems). With these tools, we proved also that a pure-state AES is robust with respect to the mixture with noise, thus providing an example of an AES of mixed states.

A few technical questions and several generalisations remain open: these were mentioned in the paper. More broadly, the study of absolute entangled set of states will have to be extended to multipartitions; and one can consider the correlative definition of absolutely entangled quantum measurements.}


\textit{Acknowledgements ---}
We thank Alastair Abbott, Shuming Cheng and Rotem Arnon-Friedmann for useful discussions. This research is supported by the National Research Foundation and the Ministry of Education, Singapore, under the Research Centres of Excellence programme. We acknowledge financial support from the Swiss National Science Foundation (Starting grant DIAQ and NCCR SwissMap).

\bibliography{myref}

\newpage
\appendix

\section{Proof of the example of absolutely entangled set}

In this section we prove that for generic bipartition $\bbC^d=\bbC^{d_1} \otimes \bbC^{d_2}$, the set of $K\equiv d_1+d_2$ state 
\ba\label{apeq1}
\begin{array}{ccl}
    \ket{\phi_1} &=& \ket{\xi_1}, \\
    \ket{\phi_k} &=& c_{k1}\ket{\xi_{1}}+c_{kk}\ket{\xi_{k}},\;k=2,...,K
    \end{array}
\ea
where $\{\ket{\xi_i}\}_{i=1,...,d}$ is an orthonormal basis of $\bbC^d$, is absolutely entangled when all $|c_{k1}| = a \in (\sqrt{\frac{(d_1-1)(d_2-1)}{d_1d_2}},1)$.
Consider a certain bipartition $\bbC^d=\bbC^{d_1} \otimes \bbC^{d_2}$, any pure state $\ket{\phi_i}$ can be written with the product of local computational basis as

\begin{equation}
\ket{\psi}=a_{1}\ket{11}+a_{2}\ket{12}+...a_{d}\ket{d_1d_2}.
\end{equation}
A necessary and sufficient condition for separablity of $\ket{\psi}$ is

\begin{equation}\label{apeq2}
a_{1}\cdot a_{n\cdot d_2+j}=a_{n\cdot d_2+1}\cdot a_{j},
\end{equation}
where integers $n\in[1,d_1-1]$, $j\in[2,d_2]$. Take $|\cdot|^2$ of both sides of Eq.~\eqref{apeq2} and summing over $n$ and $j$ we obtain 
\begin{equation}\label{apeq3}
|a_{1}|^{2}\cdot (\sum_{n=1}^{d_1-1}\sum_{j=2}^{d_{2}}|a_{n\cdot d_{2}+j)}|^{2})=\sum_{n=1}^{d_1-1}|a_{n\cdot (d_2+1)}|^2\cdot \sum_{j=2}^{d_2}|a_{j}|^{2}.
\end{equation}

Now we prove that no unitary matrix can take the all states in the constructed set into separable states, by showing the contradiction between the necessary condition \eqref{apeq3} for separability and the unitarity of the matrix. Suppose a unitary matrix $U$ takes all states into separable form, we can always let the first transformed state $U\ket{\phi_1} = U\ket{\xi_1}$ to be \pending{$\ket{11}$}. Then the remaining transformed states $U\ket{\xi_i}$ $(i>1)$ are
\begin{equation}
U\ket{\xi_i}=b_{i2}\ket{12}+b_{i3}\ket{13}+...+b_{id}\ket{d_1d_2}.
\end{equation}
Since $\{\ket{\xi_i}\}$ is an orthogonal basis, if we take coefficients $b_{ij}$ to be the $(i-1,j-1)$-th element of a $(d-1,d-1)$ matrix, the matrix would also be an unitary matrix, which is a submatrix of $U$ under the computational product basis representation. We will denote this submatrix of $U$ by $U_{s}$ and let all $|c_{k1}|=a$ hereinafter. Applying separability condition \eqref{apeq3} to the $K-1$ transformed states $U\ket{\phi_k}$, we have 
\begin{equation}\label{apeq4}
a^{2}\cdot T_{k}=(1-a^{2}) B_{k}\cdot S_k,
\end{equation}
where $S_k=\sum_{j=2}^{d_2}|b_{kj}|^{2}$, $T_{k}=\sum_{n=1}^{d_1-1}\sum_{j=2}^{d_{2}}|b_{k(n\cdot d_{2}+j)}|^{2}$, $B_{k}=\sum_{n=1}^{d_1-1} |b_{k(n\cdot d_{2}+1)}|^2$.
Since every row and column of $U_{s}$ is normalized, we have
$S_k+T_k+B_k=1$ $(k>1)$, therefore
\begin{equation}\label{apeq5}
S_k=\frac{a^{2}(1-B_k)}{B_k+a^2(1-B_k)} 
\end{equation}
Summing up Eq.~\eqref{apeq5} we have
\begin{equation}\label{apeq55}
\sum_{k=2}^{K} S_{k}=\sum_{k=2}^{K}\frac{a^{2}(1-B_k)}{B_k+a^2(1-B_k)}. 
\end{equation}
Let $B=\sum_{k=2}^{K} B_{k}$, using the unitarity of $U_{s}$ we know that $B\leq d_1-1$. Substituting one of the variables $B_{k}$ in \eqref{apeq55} (say $B_{K}$) with $B_{K}=B-\sum_{k=2}^{K-1}B_{k}$ and taking the derivatives of $\sum_{k=2}^{K} S_{k}$ with respect to every other $B_{k}$, with simple calculation we can see that the lower bound of $\sum_{k=2}^{K}S_{k}$ is attained if and only if we let $B= d_1-1$ and all $B_{k}$ equal, namely, $B_{k}=\frac{d_{1}-1}{K-1}$ for every $k$. So we have
\begin{equation}\label{apeq6}
\sum_{k=2}^{K}S_{k}\geq\frac{(K-1)(K-d_1)a^{2}}{(d_1-1)(1-a^2)+(K-1)a^2},
\end{equation}
From Eq.~\eqref{apeq6} we can also see the intuition of choosing $K=d_1+d_2$. If we choose an $a$ very close to 1, then using Eq.~\eqref{apeq6} we see that $\sum_{k=2}^{K}S_{k}$ would be close to $K-d_1$. On the other hand, since $\sum_{k=2}^{K}S_{k}$ sums up the squared norm of part of the elements in $d_2-1$ column of $U_{s}$, we have
\begin{equation}\label{apeq8}
\sum_{k=2}^{K}S_{k}\leq d_2-1,
\end{equation}
therefore if $K>d_1+d_2-1$ we can find contradiction. Now we calculate the range of $a$ of having AE. From Eq.~\eqref{apeq6} and \eqref{apeq8} we know that a necessary condition for not having AE is 
\begin{equation}\label{apeq9}
 d_2-1\geq\frac{(K-1)(K-d_1)a^{2}}{(d_1-1)(1-a^2)+(K-1)a^2},
\end{equation}
inserting $K=d_1+d_2$ and with simple calculation we obtain
:
\begin{equation}\label{apeq10}
a^{2}\leq \frac{(d_1-1)(d_2-1)}{d_1d_2}.
\end{equation}
Eq.~\eqref{apeq10} shows that if we let $a^{2}>\frac{(d_1-1)(d_2-1)}{d_1d_2}$, then $\sum_{k=2}^{K}S_{k}>d_2-1$, which violates the unitarity condition. Therefore when $a>\sqrt{\frac{(d_1-1)(d_2-1)}{d_1d_2}}$, there is no unitary that can take all the $K$ states into separable form, the set of states is absolutely entangled.

\label{app:proof}

 \section{Constructive proof of separability}
 \label{sec:sep}

Here we give a constructive proof that it is possible to turn the input set
\begin{align}
\begin{split}
    \ket{\phi_1} &= \ket{\xi_1}, \\
    \ket{\phi_2} &= c_{21}\ket{\xi_{1}}+c_{22}\ket{\xi_{2}}, \\
    \ket{\phi_3} &= c_{31}\ket{\xi_{1}}+c_{33}\ket{\xi_{3}}, \\
    \ket{\phi_4} &= c_{41}\ket{\xi_{1}}+c_{44}\ket{\xi_{4}},
\end{split}
\end{align}
separable by a general unitary, when one of the coefficients $c_{j1}$ is in $(0, 0.5]$.\\

We let $|c_{21}|\in(0, 0.5]$ and $|c_{31}|=|c_{41}|=\frac{1}{2 |c_{21}|+1}\in [0.5, 1)$, and show that we can construct a unitary $U$ that take $\ket{\xi_1}$ to $\ket{00}$ and
\begin{align}
U|\xi_{2}\rangle=b_{22}|01\rangle+b_{23}|10\rangle+b_{24}|11\rangle,\nonumber\\
U|\xi_{3}\rangle=b_{32}|01\rangle+b_{33}|10\rangle+b_{34}|11\rangle,\\
U|\xi_{4}\rangle=b_{42}|01\rangle+b_{43}|10\rangle+b_{44}|11\rangle.\nonumber
\end{align}
such that
\begin{align}\label{Coefmatr}
U_{b}=\left(
 \begin{matrix}
   b_{22} & b_{23} & b_{24} \\
   b_{32} & b_{33} & b_{34} \\
   b_{42} & b_{43} & b_{44}
  \end{matrix}
  \right)
  \end{align}
is unitary and
\begin{equation}\label{sepcon}
c_{i1}b_{i4}=c_{ii}b_{i2}b_{i3}
\end{equation}
for $i=2,3,4$, and therefore the four states can be taken separable by $U$.\\

We know that with fixed $c_{i1}$ and $c_{ii}$, the largest value of $|b_{i4}|_{max}$ attainable under separability condition (\ref{sepcon}) is attained when $|b_{i2}|=|b_{i3}|$, and we can obtain the one-to-one correspondence between $|c_{i1}|$ and $|b_{i4}|_{max}$ as
\begin{equation}
|b_{i4}|_{max}^{2}=\frac{2}{|c_{i1}|+1}-1.
\end{equation}
By inspection, we can see that the input set we construct satisfies
\begin{equation}
\sum_{i}\frac{2}{|c_{i1}|+1}=4,
\end{equation}
indicating that
\begin{equation}
\sum_{i}|b_{i4}|_{max}^{2}=1.
\end{equation}
Now we can see that if and only if $|b_{i4}|=|b_{i4}|_{max}$, the sum of the squared norm of elements in every row (column) of matrix (\ref{Coefmatr}) is 1 (a necessary condition of unitarity) with our input set. Also we have
\begin{equation}\label{sepmod}
|c_{i1}b_{i4}|=|c_{ii}b_{i2}b_{i3}|
\end{equation}
holds for $i=2,3,4$.

\indent As the modulus of the elements in matrix (\ref{Coefmatr}) have been determined, now it is left to show that we can always assign proper phases to each elements of matrix
$\left(
 \begin{matrix}
   |b_{22}| & |b_{23}| & |b_{24}| \\
   |b_{32}| & |b_{33}| & |b_{34}| \\
   |b_{42}| & |b_{43}| & |b_{44}|
  \end{matrix}
  \right)$  to make it a unitary matrix, while keeping Eq.~(\ref{sepcon}) satisfied at the same time. With simple calculation we can see that
  \begin{equation}
  |b_{3m}\cdot b_{4m}^{*}|+|b_{3n}\cdot b_{4n}^{*}|>|b_{3l}\cdot b_{4l}^{*}|,
  \end{equation}
  for $m,n,l\in\{2,3,4\}$ and $m\neq n\neq l$. This means that we can always assign phases to let the second row vector be orthogonal to the third one. Now that we have two orthogonal vectors in a three dimensional Hilbert space, it is clear that we can have the third by assigning proper phases to the first row, to make the matrix unitary.\\
  \indent Let $U_{b}'=
\left(
 \begin{matrix}
   b_{22}' & b_{23}' & b_{24}' \\
   b_{32}' & b_{33}' & b_{34}' \\
   b_{42}' & b_{43}' & b_{44}'
  \end{matrix}
  \right)
  $ denote the unitary matrix we construct, together with Eq.~(\ref{sepmod}) we have
  \begin{equation}
c_{i1}b_{i4}'=\delta_{i}c_{ii}b_{i2}'b_{i3}',
\end{equation}
  where $\delta_{i}$ is some phase. Then we see that we can add phase $-\delta_{i}$ to row $i$ of $U_{b}'$ to let Eq.~(\ref{sepcon}) hold for every $i$ while keeping the unitarity of it, so by choosing the elements as
  \begin{equation}
  U_{bij}=-\delta_{i}U_{bij}',
  \end{equation}
 we can construct an $U_{b}$ which is unitary and satisfies Eq.~(\ref{sepcon}).\\
  \indent To conclude, for the states we give, we can always find $U$ to take them separable.

  \section{Another entanglement measurement}\label{sec:entropy}
The main text presents a quantitative measure of absolutely entangled sets constructed from negativity. Here, we study the quantity given in Eq.~\eqref{eq:Etot} with a different choice of entanglement measure, namely the entropy of entanglement $S[\rho]=-\text{Tr}(\rho_A\log\rho_A)$ where $\rho_A=\text{Tr}_B(\rho)$ is the reduced state. Contrary to the negativity, the entropy of entanglement is additive. This gives a new possible interpretation to the quantity $S[\{\rho_k\}_k]$ constructed in this way, namely as the minimum entanglement of a global state $\rho^\text{tot}=\bigotimes_k \rho_k$ comprising all states in the ensemble $\{\rho_k\}_k$, over joint unitaries of the form $U^\text{tot}=\bigotimes_k U$, i.e.
\begin{equation}\label{eq:entropy}
    S[\{\rho_k\}_k] = \underset{U\in U(d)}{\min} S(U^\text{tot} \rho^\text{tot} (U^\text{tot})^\dag).
\end{equation}

In general, entanglement measures for pure two-qubit states can be related to each other. For instance, the negativity of the state $\ket{\psi_\theta}=\cos{\theta}\ket{00} + \sin{\theta}\ket{11}$ with $\theta\in[0,\pi/4]$ is $\nty{\ket{\psi_\theta}}=\cos{\theta}\sin{\theta}$, from which we can deduce
\begin{equation}
    \theta = \arctan\left(\frac{2 \mathcal{N}}{1+\sqrt{1-4\mathcal{N}^2}}\right).
\end{equation}
Therefore, the entropy of entanglement can be expressed from the negativity for these states as $S=f(\mathcal{N})$ with
\begin{equation}
f(\mathcal{N}) = h\left(\frac{1+\sqrt{1-4\mathcal{N}^2}}{2}\right),
\end{equation}
and $h(x)$ the binary entropy function. Since $f$ is a monotonically-increasing convex function, we can further bound the absolute set entropy of entanglement directly from the absolute set negativity:
\begin{eqnarray}
S[\{\rho_k\}_k] &=& \underset{U\in U(d)}{\min} \sum_k S(U \rho_k U^\dag) \nonumber\\
 &=& \underset{U\in U(d)}{\min} \sum_k f\left(\mathcal{N}(U \rho_k U^\dag)\right) \nonumber\\
 &=& \underset{U\in U(d)}{\min} K \sum_k \frac{1}{K} f\left(\mathcal{N}(U \rho_k U^\dag)\right) \nonumber\\
 &\geq& \underset{U\in U(d)}{\min} K f\left(\sum_k \mathcal{N}(U \rho_k U^\dag)/K\right) \nonumber\\
 &=& K f\left(\underset{U\in U(d)}{\min} \sum_k \mathcal{N}(U \rho_k U^\dag)/K\right) \nonumber\\
 &=& K f\left(\nty{\{\rho_k\}_k}/K\right).\label{eq:SN}
\end{eqnarray}

\begin{figure}[tb]
    \centering
    \includegraphics[scale=0.9,trim=0 1cm 0 1.5cm]{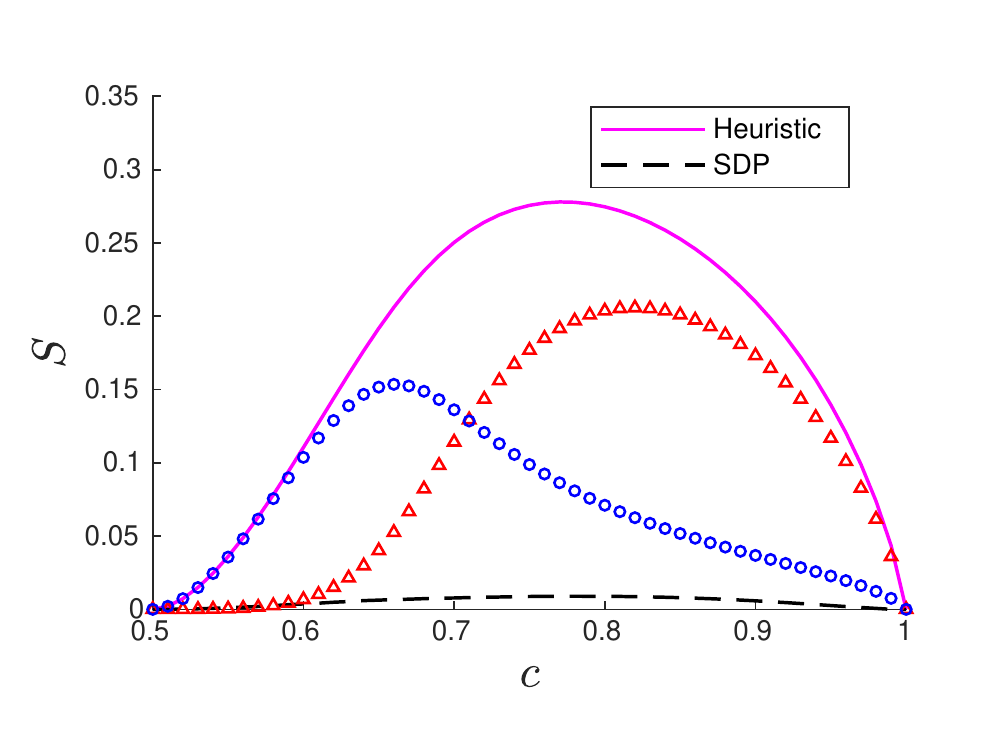}
    \caption{Absolute set entanglement entropy Eq.\eqref{eq:entropy} for the one parameter set of states Eqs.\eqref{set4} with $c_{21} = c_{31} = c_{41} = c$. The heuristic upper bound (magenta line) is computed by \texttt{fminunc}; the lower bound (black dashed) is computed via \eqref{eq:SN} and the SDP lower-bound on negativity. Albeit small the lower bound is nonzero for $c \in (0.5,1)$ since $f(\mathcal{N})>0$ when $\mathcal{N}>0$. (Blue circle) Entanglement entropy in the first state; (red triangle) Entanglement entropy in the other three states, which is equally distributed in the three states. } 
    \label{fig:minEnt}
    \end{figure}

Fig.~\ref{fig:minEnt} shows an upper bounds on the absolute set entropy of entanglement we obtain for the family of states described in~\eqref{set4} with $c_{21} = c_{31} = c_{41} = c$. This bound is obtained with a heuristic optimization as discussed in the main text. Contrary to the case of negativity, we observe that the entropy of entanglement is distributed among the four states, equally among the three symmetric states. This figure also shows a lower bound computed from the SDP lower bound on the absolute set negativity through Eq.~\eqref{eq:SN}.


\section{Convex optimization over unitaries}
\label{sec:convex}
In this appendix, we present a hierarchy of convex optimization problems that relax the non-convex optimization over unitary operators. We also discuss the localization of semi-definite constraints. Finally, we apply these methods to define a hierarchy of SDP relaxing optimization Eq.~\eqref{eq:unitaryOpt} of the main text.

\subsection{Lasserre relaxation for unitary optimization}
Let $U\in\mathcal{U}(d)$ be a unitary acting on a Hilbert space of dimension $d\geq 1$. A generic unitary optimization can be written as
\begin{align}
    \min_{U\in \mathcal{U}(d)}  & \quad f(U) \\
    \text{s.t. } 
    & \quad g_k(U) \geq 0\ \ \ \forall k\in K\nonumber
\end{align}
where $f, g_k:\mathcal{L}(\bbC^d) \to \mathbb{R}$ are forms. We are interested in the case where the these functions are polynomials in the components of $U$ with a finite degree. To capture the unitarity constraint, we choose a parametrization for $U$.

One way to parametrize a unitary operator in $\mathcal{L}(\bbC^d)$ is to write $U=\exp(H)$ for a hermitian matrix $H\in\mathcal{L}(\bbC^d)$. This only requires in total $d^2$ real scalar parameters, but involves the exponential operator. In order to avoid exponentiation, we rather parametrize $U$ as a $d \times d$ complex matrix with components $u_{i,j}\in\bbC$~\footnote{The theory of exponential cones is not as advanced as that of semi-definite cones.}. This requires $2d^2$ real parameters, i.e. twice more than necessary, but avoids exponentiation. The unitary condition then corresponds to a set of quadratic constraints on the components $u_{i,j}$:
\begin{align}\label{eq:polynomialU}
    \min_{u_{i,j} \in \bbC} &\quad f(\{u_{i,j}\}) \\
\text{s.t.}& \quad g_k(\{u_{i,j}\}) \geq 0&\forall& k \nonumber\\
& \quad \sum_m u^*_{m,i} u_{m,j} = \delta_{i,j}&\forall& 1\leq i,j\leq d \nonumber\\
& \quad \sum_m u_{i,m} u^*_{j,m} = \delta_{i,j}&\forall& 1\leq i,j\leq d. \nonumber
\end{align}
This is a special instance of polynomial optimization with polynomial objective functions and constraints. Such problems can be relaxed to a hierarchy of SDPs which converges to the optimal solution via the so-called Lasserre relaxation of polynomial optimization~\cite{lasserre2001global}. For completeness, we briefly describe it here in the simple case where the objective function $f$ is linear in the variables and the optimization involves only one nonlinear constraint $g$ (optimization~\eqref{eq:polynomialU} includes $|K|+8d^2$ real inequality constraints).

First, consider the following unconstrained optimization with $n$ real variables collected in the vector $\bx = (x_1,x_2,\ldots,x_n)$ and polynomial $p$ as objective function:
\begin{align}\label{eq:PO}
    p^*:= \min_{\bx\in \mathbb{R}^n} &\quad p(\bx).
\end{align}
For short, we denote monomials in $\bx$ as $x^{\balpha}=\prod_{i=1}^n x_i^{\alpha_i}$, where $\alpha_i$ denotes the power of variable $x_i$ (in particular $\alpha_i=0$ if $x_i$ does not appear in $x^\balpha$) and $\balpha=\alpha_1,\alpha_2,\ldots,\alpha_n\in\mathbb{N}^n$ (for conciseness, we ignore the parentheses in $\balpha$). Since any polynomial can be expressed as a linear combination of monomials, the objective function can be written $p(\bx)=\sum_\balpha p_\balpha x^\balpha$, with coefficients $\{p_\balpha\}$.

Considering now a probability measure $\mu(\bx)$ on $\mathbb{R}^n$, we write the moment associated to each monomial $y_\balpha=y_{\alpha_1,\alpha_2,\ldots,\alpha_n}=\int d\mu(\bx)x^\balpha$. For simplicity, we express averaging over the measure $\mu$ for arbitrary polynomial as a function application by $y$. For example, $y(1)=\int d\mu(\bx) =1=y_{0,0,\ldots,0}$, $y(x_1)=\int d\mu(\bx) x_1=y_{1,0,\ldots,0}$, $y(x_1^2 x_2)=\int d\mu(\bx) x_1^2 x_2=y_{2,1,0,\ldots,0}$, $y(p(x))=\sum_\balpha p_\balpha y_\balpha$, etc.

The moments $y_\balpha$ satisfy specific constraints. To see this, consider a matrix of moments $M(\mathcal{X}) = \int d \mu(x) \mathcal{X}^T \mathcal{X}  = y(\mathcal{X}^T \mathcal{X})$ defined for a set of monomials $\mathcal{X}$ (here we regard a set of monomials $\mathcal{X}$ as a row vector and $(\cdot)^T$ denotes transposition). In particular, let $\mathcal{X}_m$ be the set containing all monomials of degree up to $m\geq 0$. These sets form a hierarchy: $\mathcal{X}_0 = \{1\}$, $\mathcal{X}_1 = \mathcal{X}_0 \cup \{x_i\}_i$, $\mathcal{X}_2 = \mathcal{X}_1 \cup \{x_ix_j\}_{i,j}$ and so on. For $n=2$ variables and level $m=2$ the moment matrix takes the form
\begin{align}
    M(\mathcal{X}_2) =  \left[
    \begin{array}{c|c c | c c c}
         1       & y_{1,0} & y_{0,1}  & y_{2,0} & y_{1,1} & y_{0,2}  \\
         \hline
         y_{1,0} & y_{2,0} & y_{1,1}  & y_{3,0} & y_{2,1} & y_{1,2} \\
         y_{0,1} & y_{1,1} & y_{0,2}  & y_{2,1} & y_{1,2} & y_{0,3} \\
         \hline
         y_{2,0} & y_{3,0} & y_{2,1}  & y_{4,0} & y_{3,1} & y_{2,2} \\
         y_{1,1} & y_{2,1} & y_{1,2}  & y_{3,1} & y_{2,2} & y_{1,3} \\
         y_{0,2} & y_{1,2} & y_{0,3}  & y_{2,2} & y_{1,3} & y_{0,4}
    \end{array} \right].
\end{align}
For every set of monomials $\mathcal{X}$ this matrix is positive semidefinite, hence constraining the possible values of the moments $y_\balpha$: $\bra{v} M \ket{v} = \int d\mu(\bx)(\bra{v} \mathcal{X}^T) (\mathcal{X} \ket{v})= \int d\mu(\bx) ||\mathcal{X} \ket{v}||^2\geq 0$ for all $\ket{v}$.

Seeing now the moments $y_\balpha$ as variables, we can define a hierarchy of semi-definite program indexed by $m$:
\begin{align}\label{eq:PO2}
    q^* = \min_{\by} &\quad \sum_\balpha p_\balpha y_\balpha \\
    \text{s.t.} &\quad M(\mathcal{X}_m) \geq 0, \nonumber
\end{align}
where $\by$ is the vector containing all moments $\{y_\balpha\}$. If $\bx^*$ is a solution to the original minimization~\eqref{eq:PO} with optimal value $p^* = p(\bx^*)$, then the Dirac distribution $\mu(\bx) = \delta(\bx^*)$ gives rise to a valid variable assignment for the new optimization~\eqref{eq:PO2} yielding an identical value for the objective function: $\sum_\balpha p_\balpha y_\balpha=p^*$. Therefore we must have $q^*\leq p^*$, i.e.~\eqref{eq:PO2} is a relaxation of~\eqref{eq:PO}. The hierarchy was shown to converge, that is $q^* = p^*$ when $m$ tend to infinity.

Now we turn to the constrained case. If the optimization is also subjected to polynomial constraints, such as
\begin{align}\label{eq:POC}
    p^*:= \min_{\bx\in \mathbb{R}^n} &\quad p(\bx)\\
    \text{s.t.}&\quad g(\bx)\geq0, \nonumber
\end{align}
each constraint must be localized with respect to a monomial basis $\mathcal{X}$. This is achieved by constructing the moment matrix $M_{g}(\mathcal{X}) = \int d\mu(\bx) \mathcal{X}^T \mathcal{X} g(\bx)$. For example, if we localize $x_1+x_2^2 \geq 0$ by $\mathcal{X} = \{1,x_1,x_2\}$, we have
\begin{align}
    M_{x_1+x_2^2}(\mathcal{X}) =  \left[
    \begin{array}{c|c c}
         y_{1,0}+y_{0,2} & y_{2,0}+y_{1,2} & y_{1,1}+y_{0,3} \\
         \hline
         y_{2,0}+y_{1,2} & y_{3,0}+y_{2,2} & y_{2,1}+y_{1,3} \\
         y_{1,1}+y_{0,3} & y_{2,1}+y_{1,3} & y_{1,2}+y_{0,4}
    \end{array} \right].
\end{align}

Similarly to the case of the moment matrix $M(\mathcal{X})$, the matrix $M_{g}(\mathcal{X})$ is semi-definite positive, which gives rise to a constraint that can be imposed when the moments $y_\balpha$ are seen as variables. Indeed, $\bra{v} M_{q(\bx)} \ket{v} = \int d\mu(\bx) q(\bx) (\bra{v} \mathcal{X}^T) (\mathcal{X} \ket{v})= \int d\mu(\bx) q(\bx) ||\mathcal{X} \ket{v}||^2\geq 0$ for all $\ket{v}$. Therefore, optimization Eq.~\eqref{eq:POC} is relaxed to
\begin{align}\label{eq:PO2}
    p^* \geq \min_{\by} &\quad \sum_\balpha p_\balpha y_\balpha \\
    \text{s.t.} &\quad M(\mathcal{X}^1) \geq 0, \nonumber\\
    &\quad M_g(\mathcal{X}^2) \geq 0, \nonumber
\end{align}
for any set of monomials $\mathcal{X}^1$, $\mathcal{X}^2$.

\subsection{Localizing semi-definite positive constraints}
Coming back to optimization~\eqref{eq:unitaryOpt} of the main text, we see that we are now able to address its unitarity constraint. Namely, choosing a suitable parametrization of $U$, the unitarity constraint takes the form of a polynomial constraint which can be relaxed with the methods of polynomial optimization as described above. 
\change{Our optimization~\eqref{eq:unitaryOptElement} also involves semi-definite constraints on the density matrices $\sigma_k^\pm$. We thus need to localize these constraints as well.} 
While localizing matrices have been described for scalar equality and inequality constraints (and therefore for element-wise constraints on matrices elements) we are not aware of a description of localizing matrices for semi-definite constraints. We describe this now.

Denoting again by $\bx$ the set of scalar variables, we consider a matrix $G(\bx)$ with elements $g_{i,j}(\bx)$ polynomials in $\bx$. To localize the semi-definite constraint $G(\bx)\geq 0$, we consider again a basis of monomials $\mathcal{X}$. The localizing matrix then reads
\begin{align}
     M_{G}(\mathcal{X}) = \int d\mu(\bx) G(\bx) \otimes \mathcal{X}^T \mathcal{X}.
\end{align}
Since the Kronecker product of two PSD matrices is PSD, $G(\bx) \otimes \mathcal{X}^T \mathcal{X} \geq 0$. Then by convexity, we must have $M_{G}(\mathcal{X}) \geq 0$. Thus semi-definite positive constraints $G(\bx)\geq 0$ is relaxed to $M_{G(x)}(\mathcal{X}) \geq 0$.

As an example, the constraint 
    $G=\begin{bmatrix}
        x_1 & x_2 \\
        x_2 & x_3 
    \end{bmatrix} \geq 0$
localized by $\mathcal{X}=\{1, x_1, x_2, x_3\}$ is given by the matrix
\begin{eqnarray}
    &&M_{G}(\mathcal{X})=\\
    &&\left[
    \begin{array}{cc|cc|cc|cc}
        y_{1,0,0} & y_{0,1,0} & y_{2,0,0} & y_{1,1,0} & y_{1,1,0} & y_{0,2,0} & y_{1,0,1} & y_{0,1,1}\\
        y_{0,1,0} & y_{0,0,1} & y_{1,1,0} & y_{1,0,1} & y_{0,2,0} & y_{0,1,1} & y_{0,1,1} & y_{0,0,2}\\
        \hline
        y_{2,0,0} & y_{1,1,0} & y_{3,0,0} & y_{2,1,0} & y_{2,1,0} & y_{1,2,0} & y_{2,0,1} & y_{1,1,1}\\
        y_{1,1,0} & y_{1,0,1} & y_{2,1,0} & y_{2,0,1} & y_{1,2,0} & y_{1,1,1} & y_{1,1,1} & y_{1,0,2}\\
        \hline
        y_{1,1,0} & y_{0,2,0} & y_{2,1,0} & y_{1,2,0} & y_{1,2,0} & y_{0,3,0} & y_{1,1,1} & y_{0,2,1}\\
        y_{0,2,0} & y_{0,1,1} & y_{1,2,0} & y_{1,1,1} & y_{0,3,0} & y_{0,2,1} & y_{0,2,1} & y_{0,1,2}\\
        \hline
        y_{1,0,1} & y_{0,1,1} & y_{2,0,1} & y_{1,1,1} & y_{1,1,1} & y_{0,2,1} & y_{1,0,2} & y_{0,1,2}\\
        y_{0,1,1} & y_{0,0,2} & y_{1,1,1} & y_{1,0,2} & y_{0,2,1} & y_{0,1,2} & y_{0,1,2} & y_{0,0,3}
    \end{array} \right].\nonumber
\end{eqnarray}

Note that this construction of localizing matrices for semi-definite constraints generalizes directly to non-commutative polynomial optimization~\cite{Pironio2010}. In particular, the hierarchy presented in~\cite{Wang2019} can be understood as a relaxation of the Gram matrix constraint in a non-commutative algebra.

\subsection{Full optimization}
We can now formulate the full SDP relaxation of our original problem~\eqref{eq:unitaryOpt}.

Let us start by defining all scalar variables appearing in Eq.~\eqref{eq:unitaryOptElement}. After separating complex numbers into real and imaginary components, we parametrize 4-by-4 complex $U$ with 32 real variables $(u_a)_{a\in[32]}$, and each $\sigma_k^\pm$ with 16 variables $(\xi_{k,s,b})_{b\in[16]}$, where $k\in[4]$ and $s\in\{\pm\}$. Optimization~\eqref{eq:unitaryOptElement} thus involves in total $n=160$ variables $x_i$, $i=1,\ldots,n$.

As discussed above, our SDP relaxation is defined on variables $y_\balpha$ corresponding to moments involving the variables in $\bx$. Only moments appearing in some constraints are taken into account, and all corresponding variables are considered as free variables.

To limit the number of constraints in the optimization, each constraint of Eq.~\eqref{eq:unitaryOptElement} is localized with a specific set of monomials. For conciseness, we write $\mathcal{X}=\{u^r\}$ for the set of all monomials appearing in powers of $u$ up to $r$, e.g. $\{u\}=\{1\}\cup\{u_i\}_i$ and $\{u^2\}=\{1\}\cup\{u_i\}_i\cup\{u_i u_{i'}\}_{i,i'}$. Also, we do not include the constraint $U^\dagger U=\Id$ since it doesn't seem to be very helpful. The SDP then reads:

\begin{align}
\begin{split}
    \nty{ \{ \rho_k \} } \geq \min_{y_\balpha} & \quad \sum_k \Tr[\sigma^-_k] \\
    \text{s.t.}
    & \quad M(\{u^2\}) \geq 0, \\
    & \quad M(\{(\xi_{k,\pm})^2\}) \geq 0,\\
    & \quad M_{(\sigma^\pm_k)^{T_A}}(\{u\}) \geq 0, \\
    & \quad M_{U\rho_kU^\dagger-\sigma_k^++\sigma_k^-}(\{u\})=0,\\
    & \quad M_{UU^\dagger-\Id}(\{u, \xi_{k}\})=0 \\
\end{split}
\end{align}

\end{document}